\documentclass{article}

\usepackage[final]{neurips_2022}
\usepackage[dvipsnames]{xcolor}

\usepackage[utf8]{inputenc} 
\usepackage[T1]{fontenc}    
\usepackage{hyperref}       
\usepackage{url}            
\usepackage{booktabs}       
\usepackage{amsfonts}       
\usepackage{nicefrac}       
\usepackage{microtype}     
\usepackage{xcolor}        
\usepackage{graphicx}
\usepackage{caption}
\usepackage{subcaption}

\title{Towards Foundation Models for Relational Databases [Vision Paper]}

\author{%
  Liane Vogel \\
  Technical University of Darmstadt\\
  \And
  Benjamin Hilprecht \\
  Technical University of Darmstadt\\
  \AND
  Carsten Binnig \\
  Technical University of Darmstadt \& DFKI \\
}

\begin{document}

\maketitle

\begin{abstract}\vspace{-1.5ex}Tabular representation learning has recently gained a lot of attention.
However, existing approaches only learn a representation from a single table, and thus ignore the potential to learn from the full structure of relational databases, including neighboring tables that can contain important information for a contextualized representation.
Moreover, current models are significantly limited in scale, which prevents that they learn from large databases. 
In this paper, we thus introduce our vision of relational representation learning, that can not only learn from the full relational structure, but also can scale to larger database sizes that are commonly found in real-world.
Moreover, we also discuss opportunities and challenges we see along the way to enable this vision and present initial very promising results. 
Overall, we argue that this direction can lead to \emph{foundation models for relational databases} that are today only available for text and images. 
\vspace{-3ex}
\end{abstract}

\vspace{-1.5ex}
\section{Introduction}
\vspace{-2ex}\paragraph{Motivation.}  Tabular representation learning has recently gained a lot of attention.
For example, approaches like RPT \citep{tang2021rpt} or TURL \citep{deng2020turl} have shown  that they can provide deep contextualized representations by self-supervised pre-training that enable many data engineering tasks such as data cleaning, entity resolution, column type annotation etc. to be efficiently derived from the representation with low overhead. \\
However, these existing approaches only learn a representation from a single table and thus ignore the potential to learn a representation from the full structure of relational databases, i.e., by taking neighboring tables as well as relationships between tables into account.
We argue that being able to process signals from the full relational structure such as neighboring tables enables us to utilize valuable information in relational databases. 
For example if the information that a neighboring table is called \emph{actors} is available, it is easier for a model to infer that the missing table name of the table under consideration is more likely to be \emph{movies} than \emph{music albums}.\\
Moreover, many of the recent approaches like TURL or RPT significantly build on language models (LMs) that have been pre-trained on large textual corpora. It was even shown that LMs not adapted for specific data engineering tasks such as entity resolution can already achieve state-of-the-art performance \cite{https://doi.org/10.48550/arxiv.2205.09911}.
Unfortunately, while these approaches based on pre-trained LMs have shown remarkable results and outperform other approaches such as EmbDI or Termite \citep{cappuzzo2020embdi, fernandez2019termite}, they do not scale to large data sizes, since they are often limited by the restrictive input sizes of LMs. 
While this limitation is negligible for the scenarios that these approaches have been originally developed for (e.g., for web tables or tabular data in CSV files), real-world relational databases, as they can be found in enterprises today, are typically much larger in size \citep{krueger2011database}.\\
Furthermore, there has been some work on representation learning for knowledge graphs \citep{DBLP:journals/access/AlamRNMAVL22, DBLP:journals/tacl/WangGZZLLT21}. However, this line of work typically focuses on downstream tasks such as link prediction, which are very different from the tasks we are focusing on. 
\vspace{-2ex}\paragraph{Contributions.}
In this paper, we thus introduce our vision of relational representation learning that can not only learn from the full relational structure instead of only single tables, but also scales to large databases.
The main idea of the new model architecture to tackle these issues is to combine language models (LMs) with graph neural networks (GNNs), which is a combination that has already successfully been used in other domains \citep{ioannidis2022efficient}.
As the core contribution of this paper, we discuss the design of such a model for relational representation learning.
The main idea behind the model is that we use LMs to encode individual table rows and their schema to model the relational structure within tables (e.g., which rows belong to the same table) as well as structure across tables (e.g., relationships between rows).\\
Moreover, as a second contribution, we show initial highly promising results of using our new model architecture for three representative tasks on two data sets, where we \textbf{outperform existing single-table models by more than $2\times$ in accuracy} in the best case.
Overall, based on our initial results, we believe that our approach for learning the representation of relational data can actually lead us towards \textbf{foundation models for relational data} which exist today only for text \& images.

\begin{figure}
    \centering
    \includegraphics[width=0.95\linewidth]{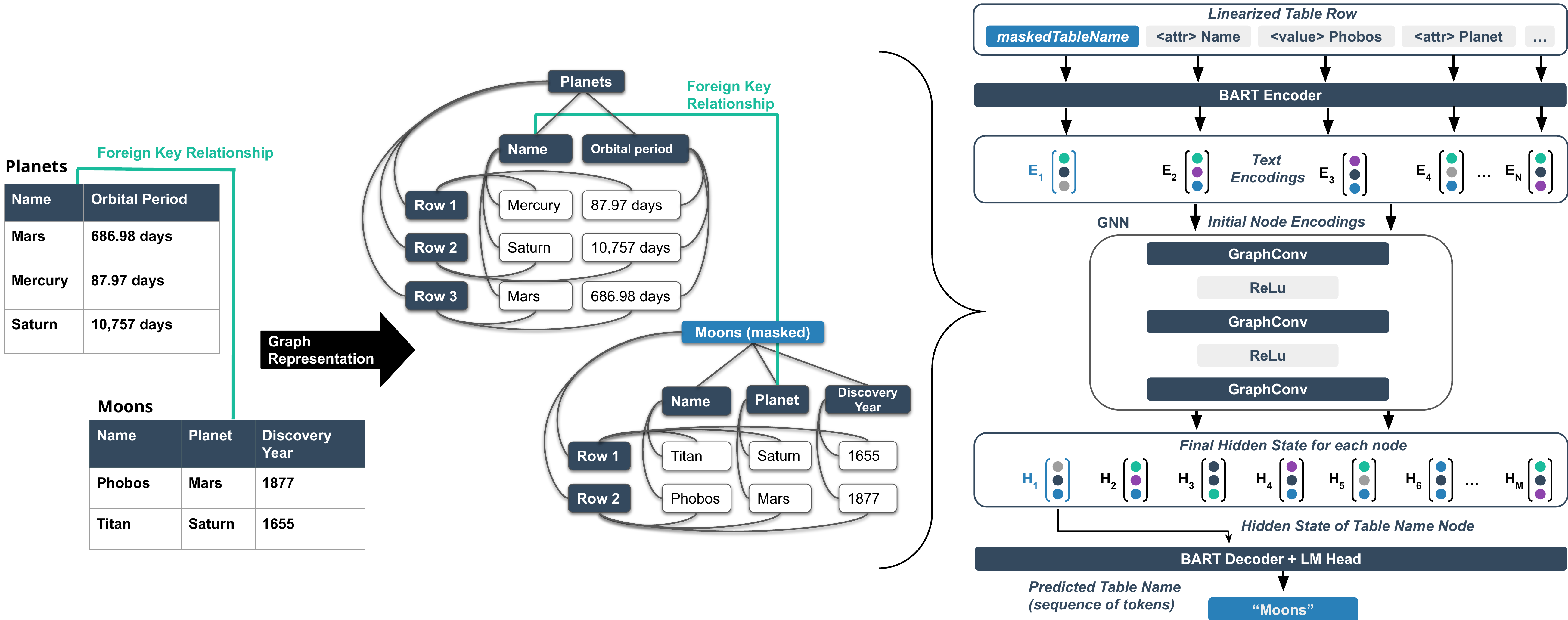}
    \caption{Relational Foundation Models combine Language Models (LMs) with Graph Neural Networks (GNNs). An example of the graph representation is shown on the left-hand side and the architecture is shown on the right-hand side.}
    \vspace{-3.5ex}
    \label{fig:foundation-model}
\end{figure}

\vspace{-1.5ex}
\section{Overall Vision}
\vspace{-1.5ex}
Foundation models like GPT-3 \citep{brown2020gpt3} for text are pre-trained in a self-supervised manner on very large corpora and can be adapted to solve downstream tasks with comparatively little effort. 
However, for relational data, a foundation model still needs to be developed. 
\vspace{-2ex}\paragraph{Relational Foundation Models.} The vision that we propose in this paper is to pre-train a high-capacity model on a large corpus of different relational databases. The data is supposed to cover various sizes of databases from a large variety of domains. This allows for a model that generalizes out-of-the-box to previously unseen databases.  
As we have shown in our initial experiments in Section \ref{sec:results}, such a relational foundation model has the potential to greatly facilitate the automation of data engineering tasks where 
a small amount of annotated training data suffices to adapt the model to tasks like schema matching or entity resolution.\\
To enable such a pre-trained relational model that can learn from a wide range of different relational databases in a self-supervised manner, we propose a new model architecture as shown in Figure~\ref{fig:foundation-model}.
As discussed before, our proposed model architecture for relational representation learning is a combination of language models (LMs) with graph neural networks (GNNs). 
However, blindly combining LMs with GNNs will not lead to a foundation model for relational data that can provide high performance and scale to large databases.

\vspace{-2ex}\paragraph{Core Design Principles.} 
In the following, we present the core design principles of our model architecture to achieve scalability and high accuracy of learned relational representations.\\
\emph{Design for Scale.} Existing approaches such as TURL \citep{deng2020turl} are limited in their scalability since they linearize the full table (i.e., all rows) as input to a language model.
To address this issue, in our model architecture we only linearize individual rows together with their schema information (i.e., column names and table names) which we use as input for an LM encoder (i.e., BART \citep{lewis2020bart} in our case) as shown in Figure \ref{fig:foundation-model} (right).
The row-wise embeddings are then combined by the GNN to learn a relational embedding across rows and tables as we discuss next.
The representations learned by our current model implementation can be used for enabling various downstream tasks by using a task-specific LM-head as shown in Figure \ref{fig:foundation-model} (right). 
Similar to our model, RPT also encodes table data row-wise, but as a major difference RPT does not learn higher-level representations on the table-level or even across tables, and thus the representations lead to much lower accuracy for various tasks, as we show in our initial evaluation in Section \ref{sec:results}. \\
\emph{Design for Expressiveness.} In order to learn a relational representation on the level of tables and across tables, we model the structure of the relational data as a graph that we use as basis for the GNN.
Modeling relational data as graph for a GNN comes natural in many places and various different approaches exist. In fact, our graph structure as shown in Figure \ref{fig:foundation-model} (right) is inspired by \cite{cvitkovic2020databasegnn}.
However, it is far from obvious how LMs and GNNs should be best combined to learn a deep contextualized representation of relational data and how such models can be efficiently trained. 
In the following, we hence discuss the current state of our implementation as well as open challenges.

\vspace{-1.5ex}
\section{Current State and Open Challenges}
\label{sec:opportunities}
\vspace{-1.5ex}

In the following, we discuss the current state of our model implementation as well as open challenges. 

\vspace{-2.5ex}\paragraph{Pre-training Procedure.}
With our model, we have the ability to learn representations by taking the full relational structure into consideration.
For example, to learn a representation of the table \emph{Moons} in Figure \ref{fig:foundation-model}, all columns of that table (with the names and cell values) as well as neighboring tables are considered.
As a pre-training task to learn a relational representation, we use masked value reconstruction on the different levels of the model; i.e., we use masked cell values, masked column names, and masked table names.
In our current implementation, we first fine-tune a pre-trained LM on individual table rows and then freeze its weights when pre-training the representations (i.e., for cell values, columns and tables) using the GNN, which propagates the information across rows and tables based on message passing. 
Another alternative would be to pre-train the entire model including the LM end-to-end, which is clearly more resource intensive but might yield even better accuracy. 

\vspace{-2.5ex}\paragraph{Datasets for Pre-training.}
For pre-training, a large training corpus of relational databases with different data characteristics is needed.
Unfortunately, the available large table corpora such as \emph{gitTables} \citep{hulsebos2021gittables} contain only single (web-)tables and thus lack relationships between the tables.  
As such, constructing a corpus of relational databases with connected tables is an open challenge.
One starting point is the \emph{Relational Learning Repository} \citep{motl2015relationalfit}. Unfortunately, this repository only has a limited number of databases and many databases are also rather small in size. 
As such, another promising direction is to use \emph{wikiData} \citep{vrandevcic2014wikidata} to automatically extract a corpus of relational databases from separate Wiki domains with tables that are connected. 

\vspace{-2.5ex}\paragraph{Support for Wide Tables.}  
As discussed before, our model architecture has the ability to learn representations from databases with a large number of rows.
However, for wide tables with a high number of columns, as they can be found in real-world databases as well, the serialized representation of a table row might still exceed the maximum number of tokens that an LM can process (i.e. 1024 tokens for BART).
For supporting wide tables in our model architecture, several options exist, such as truncating the tables or vertically splitting them into several smaller (connected) tables. However, how to split tables in an optimal manner without loosing context in the row encoding is not clear.

\vspace{-2.5ex}\paragraph{Efficient Learning for Large Databases.}  
Relational database tables can often be very large in size and thus also lead to large graphs. 
However, processing large graphs with standard GNN techniques is costly regarding time and resources. 
One obvious way to tackle this is to use only smaller table samples or use only n-hop neighboring tables, which however, means that we lose context for learning relational representations.
An alternative, more promising direction is to train a model on the full graph with more efficient training procedures, such as specialized approaches for large graphs, e.g., GraphSAGE \citep{10.5555/3294771.3294869}.
Another approach could be to express relational data as acyclic directed graphs by performing a breadth-first traversal (BFT) of the data.  
For such acyclic directed graphs, recent work \citep{thost2021directed} has shown that a single-pass training procedure can be used that provides much more efficient training for large graphs, in contrast to using multiple rounds of message passing as common in GNNs.

\vspace{-2.5ex}\paragraph{Featurization of Relational Data.} 
In our current prototype, we only use the row-wise embeddings of the LM to initialize the nodes of the GNN. 
However, in the future we plan to also include some statistical features as input to the graph nodes, e.g., min/max values or information on distinct values of columns that could further improve the model performance (e.g., for numerical columns). 
Moreover, including some statistical features will potentially allow a foundation relational model to be used for a much wider set of tasks on relational data, such as learned approximate query answering \citep{DBLP:journals/pvldb/HilprechtSKMKB20,DBLP:conf/sigmod/MaT19} or even for learning database internal tasks, such as cardinality estimations \citep{DBLP:journals/pvldb/YangLKWDCAHKS19} that today rely on distinct model architectures.  

\vspace{-1.5ex}
\section{Initial Evaluation}
\vspace{-1.5ex}
\label{sec:results}
In this section, we present initial very promising results of our approach.
As a main result, we show that including the full relational structure is beneficial for learning the relational structure.
\vspace{-2ex}\paragraph{Pre-training of Model.} For pre-training the model, as mentioned before, we train the LM (BART in our case) and the GNN separately. 
For training BART on table data, we serialize table rows as discussed in \cite{yin2020tabert, tang2021rpt} and train BART to reconstruct masked table names, column names and cell values per row. 
As GNN, we use a graph convolutional neural network \citep{kipf2017gcn}. 
For training the GNN, we use the BART encoder to compute node embeddings and the BART decoder to convert the representation of a masked node in the GNN back into natural language text. 
We calculate the cross entropy loss between the decoder output and the true label and adapt the gradients of the GNN. 

\vspace{-2ex}\paragraph{Training Data.}
The goal of our experiment is to demonstrate that incorporating the full relational structure can improve tabular representation learning. Since to the best of our knowledge there are no baseline approaches available, which support multi-table relational datasets, we resort to single table datasets in this experiment, in particular \emph{gitTables} \citep{hulsebos2021gittables} and \emph{wikiTables} \citep{bhagavatula2015tabEL}. However, interestingly, encoding the data of single tables as graphs and thus incorporating the relational structure more efficiently already leads to significant improvements compared to RPT, which uses only a LM.
For training, we take a subset of 10,000 tables of each corpus and split it into 70/20/10 for train, validation and test set.
The checkpoint with the best accuracy on the validation set is used for evaluation and we report results as an average of three runs. 

\vspace{-2ex}\paragraph{Experiments.} In our experiments, we use a pre-trained BART model from the HuggingFace library \citep{wolf2020transformers} which we denote as BART\textsubscript{\textit{text}}. For our model, we further fine-tune this model on each of the two tabular datasets as discussed before.
We denote this model as BART\textsubscript{\textit{table}} and use it as our encoder to create initial embeddings for the GNN.
In addition, we also show two variants of our model as an ablation study, where we use the pre-trained model BART\textsubscript{\textit{text}} only as encoder or both encoder/decoder, without fine-tuning.
\vspace{-2ex}\paragraph{Initial Results.} The results of three data engineering tasks that have been used in the literature \citep{deng2020turl} are shown in Table \ref{tab:results}. 
Our results show, that we achieve considerably higher performance when compared to the RPT baseline, especially for the tasks of missing value prediction and column name detection.
RPT outperforms our approach only on the table name detection task on \textit{gitTables} data, where we had to use the filenames as table names, which are often not very expressive. Hence we speculate that learning from the full table structure does not help our approach.
Overall, our results demonstrate, that there is value in taking advantage of the relational structure in form of a graph, as it allows the model to incorporate more context.
Moreover, our ablation studies show, that it is valuable to fine-tune the LM to tabular data, as using a LM pre-trained only on text results in much lower performance. 

\begin{table}[]
\centering
\small
\caption{Results of our initial experiments on three different tasks. The RPT Baseline is our re-implementation of RPT by \cite{tang2021rpt}, since the code of RPT was not available.}
\label{tab:results}
\resizebox{\columnwidth}{!}{%
\begin{tabular}{llccc}
\toprule
Dataset    & Approach                                         & \multicolumn{3}{c}{Accuracy for mask reconstruction {[}\%{]}}                                                                                                                                                                                                 \\
\midrule
           &                                                  & \multicolumn{1}{c}{\begin{tabular}[c]{@{}c@{}}Task 1: \\ Missing Values\end{tabular}} & \multicolumn{1}{c}{\begin{tabular}[c]{@{}c@{}}Task 2: \\ Column Name Detection\end{tabular}} & \multicolumn{1}{c}{\begin{tabular}[c]{@{}c@{}}Task 3: \\ Table Name Detection\end{tabular}} \\
\midrule
wikiTables & BART\textsubscript{\textit{table} \textcolor{blue}{(encoder\&decoder)}} (RPT Baseline )              & \multicolumn{1}{c}{20.75}                                                          & \multicolumn{1}{c}{66.88}                                                           & \multicolumn{1}{c}{36.99}                                                          \\
           & BART\textsubscript{\textit{table} (encoder\&decoder)} + GNN (Ours)                         & \multicolumn{1}{c}{\textbf{46.15}}                                                 & \multicolumn{1}{c}{\textbf{83.91}}                                                  & \multicolumn{1}{c}{\textbf{37.85}}                                                 \\
           & BART\textsubscript{\textit{text} (encoder)} + BART\textsubscript{\textit{table} (decoder)} + GNN  (Ablation Study) & \multicolumn{1}{c}{37.24}                                                          & \multicolumn{1}{c}{63.77}                                                           & \multicolumn{1}{c}{37.24}                                                          \\
           & BART\textsubscript{\textit{text} (encoder\&decoder)}+ GNN (Ablation Study)                & \multicolumn{1}{c}{24.25}                                                          & \multicolumn{1}{c}{20.18}                                                           & \multicolumn{1}{c}{3.13}                                                           \\
\midrule
gitTables  & BART\textsubscript{\textit{table} (encoder\&decoder)} (RPT Baseline )              &                   21.65                                                                 &             46.63                                                                        &                   \textbf{59.71  }                                                               \\
           & BART\textsubscript{\textit{table} (encoder\&decoder)} + GNN (Ours)                         &        \textbf{ 52.63 }                                                                          &                     \textbf{90.04}                                                                &               52.63                                                                     \\
           & BART\textsubscript{\textit{text} (encoder)}+ BART\textsubscript{\textit{table} (decoder)} + GNN  (Ablation Study) &                49.00                                                                    &                                    73.15                                                 &               39.54                                                                     \\
           & BART\textsubscript{\textit{text} (encoder\&decoder)} + GNN (Ablation Study)                &         35.32                                                                           &                                          24.45                                           &                     11.78                                                           \\
\bottomrule
\end{tabular}%
}
\vspace{-3.5ex}
\end{table}

\vspace{-1.5ex}
\section{The Road Ahead}
\vspace{-2ex}
In this paper, we have described our vision of using representation learning for relational databases.
We discussed opportunities and challenges of developing foundation models for relational data.
Our initial results have shown that representing tables as a combination of GNNs and LMs leads to learning more comprehensive representations.
Overall, this a an important but only first step towards enabling the vision of foundation relational models and many open research challenges need to be addressed to enable the full vision.

\newpage
\bibliographystyle{ACM-Reference-Format}
\bibliography{references}

\end{document}